\documentclass[10pt]{article}
\usepackage{hyperref}
\usepackage{pgf, tikz}
\usepackage{pgfplots}
\usepackage{xcolor}
\usepackage{mathrsfs}  

\usepgfplotslibrary{fillbetween}

\title{Analytical model of the evolution of surfaces topography during sliding wear}

\author{No\"{e}l Bruneti\`{e}re \thanks{Institut Pprime
	CNRS, University of Poitiers - Dept GMSC - Poitiers - France - noel.brunetiere@univ-poitiers.fr}}

\begin{document}

\maketitle

\begin{abstract}
During the wear process of surfaces in sliding friction, there is a running-in period during which the topography of surfaces changes with time before reaching the steady wear regime. In the steady wear regime, the statistical parameters used to describe the topography of the surfaces remain almost constant. Some experimental studies have shown that starting friction tests with different surface finish levels leads to the same final topography of surfaces in the regime of steady wear. This paper proposes an analytical model to describe the evolution of the topography of surfaces during sliding wear. First of all, the Greenwood and Williamson approach is used to describe the contact between nominally flat rough surfaces. The asperities in contact may undergo plastic deformation or adhesion with the opposing surface. Using a plasticity criterion and an adhesion criterion, it is possible to obtain a differential equation for the evolution of the standard deviation of the asperities of the surfaces. This equation has an analytical solution that is in good agreement with experimental results from the literature. It is shown that the final surface topography is the result of the competition between abrasive wear and adhesive wear. The model is then used to describe different wear processes from polishing to galling.

\end{abstract}




\section{Introduction}

In sliding wear, the intensity of wear is generally characterized by the wear coefficient $k$. This parameter was introduced in the pioneering work of Archard in 1953 \cite{Archard1953} where it was shown that the wear rate (volume of worn material per unit sliding) is proportional to the normal force divided by the hardness $H$ of the softer material. $k$ is the proportionality coefficient defined as the probability of a junction  to generate a debris. If $k$ is constant, the wear regime is steady. However, during sliding, the wear rate is not constant, and, more particularly, there is first a running-in period where $k$ varies during the initial part of the sliding \cite{Queener1965,Bonny2010, Khonsari2021}. This change in $k$ is explained by a variation of the topography of the sliding surfaces \cite{Queener1965,Bonny2010, Khonsari2021}.

Some models have been proposed to describe the evolution of the wear rate during the running-in period. Queener et al. \cite{Queener1965} found that the wear rate evolves from its initial value to its steady state value following a decreasing exponential function. The transient wear rate amplitude is proportional to the initial arithmetic roughness height $Ra$  of the surfaces. Chandresekaran \cite{Chandrasekaran1993} proposed a better expression of the transient wear rate amplitude with a more advanced relation to the roughness height. The exponential evolution has been successfully used in several studies \cite{Kumar2002, Lijesh2019}. Hanief and Wani have shown that the wear rate is linked to $Ra$ with a power law during the running-in. More recently, Varenberg used the material ratio curve and the logistic function to calculate the wear rate evolution during the running period \cite{Varenberg2022}. In these two papers, the theoretical results were experimentally verified. The literature clearly shows that it is necessary to know the evolution of the surface topography to estimate the wear rate during the running-in period.

Considering a given wear particle size distribution, Sugimura et al. \cite{Sugimura1986} modified the height probability density function of the surface roughness with time. Knowing this, they were able to calculate the new R.M.S. roughness height, $Rq$ as well as other non-Gaussian parameters like the skewness coefficient. This work was extended by Jeng and Gao who started with initially non-Gaussian surfaces \cite{Jeng2000}. The results of these two papers show that the $Rq$ parameter decreases during the running-in period meaning that the roughness height decreases. This result has been confirmed by several experimental studies \cite{Bonny2010,PhamBa2021}. However, in the same experimental studies \cite{Bonny2010,PhamBa2021}, it is shown that if the surfaces are initially polished, the surface roughness increases during the running-in period.  This result cannot be reproduced by the model of Sugimura et al.  \cite{Sugimura1986} where it is assumed that the wear particles have a size lower than the width of the probability density function of the roughness that is truncated with time due to abrasion and plastic deformation. Thus the roughness can only decrease with time with this model. It is necessary to consider adhesive wear which can occur for very smooth surfaces and increases the roughness height of the rubbed surfaces due to wear debris formation \cite{Milanese2019}. Moreover, in the work of Pham-Ba and Molinari \cite{PhamBa2021}, the steady state roughness height reached after the running-in period is independent of the initial roughness height of the surfaces. The same result was obtained by \cite{Bonny2010} with different grades of the same material and when the initial arithmetic roughness height is lower than 1 $\mu$m. A steady state roughness height independent of the loading and initial roughness height was also obtained by Kumar et al. \cite{Kumar2002}. There exists a steady roughness height linked to the properties of the rubbed materials and the wear process \cite{Aghababaei2022}.

This paper proposes an analytical model to describe the evolution of the topography of surfaces during sliding wear. First of all, the Greenwood and Williamson approach is used to describe the contact between nominally flat rough surfaces. The asperities in contact may undergo plastic deformation or adhesion with the opposing surface. Using a plasticity criterion and an adhesion criterion, it is possible to obtain a differential equation for the evolution of the standard deviation of the asperities of the surfaces in the way proposed by Nosonovsky \cite{Nosonovsky2010}. This equation has an analytical solution that will be compared to experimental data. Then the model is applied to different cases from polishing to galling.

\section{Material and methods}

\subsection{Contact model}

\begin{figure}[h]
\centering
\includegraphics[height=2cm]{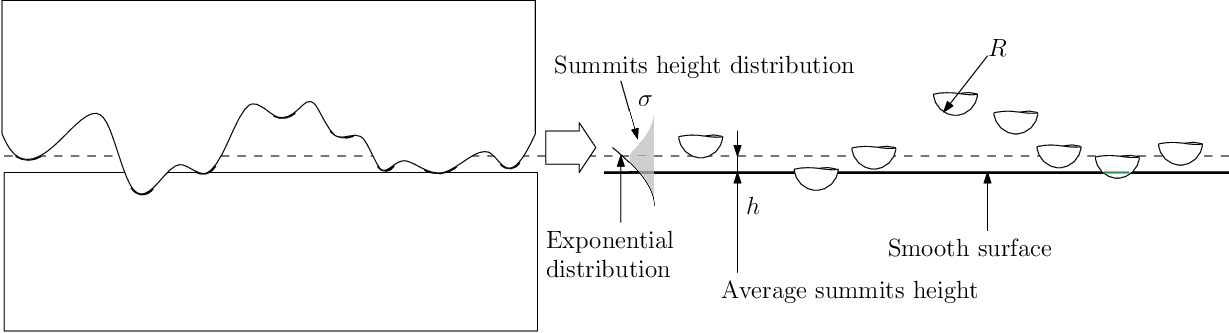}
\caption{Configuration of the contact problem }\label{fig:conf_problem}
\end{figure}

The contact between the rubbing solids is described by the Greenwood and Williamson model \cite{Greenwood1966} based on the following assumptions (see figure \ref{fig:conf_problem}). One of the surfaces is flat and smooth and the counter-body is rough. All the summits of the asperities are spherical with a uniform radius $R$. The $N$ summits have a random vertical distribution with a standard deviation $\sigma$. Each summit in contact with the contact body behaves as a spherical Hertzian contact. In addition, the vertical distribution of the summits is supposed to be exponential. This assumption is verified for the top summits of the rough surface. The elastic behavior assumption could be discussed. However, if the elastic limit is reached, plastic deformation will occur and modify the surface topography, and reduce the contact stress. This point will be discuss later. In addition, real cases where the two counter-surfaces are rough can be analyzed with the current approach if the roughness of the two surfaces are combined. However, we will focus here on the case where the flat surface is rigid.

Based on \cite{Greenwood1966} , The real contact area $A$ is :
\begin{equation}\label{eq:A}
A=\pi R N \frac{\sigma}{2} \exp\left(-\frac{2h}{\sigma} \right)
\end{equation}
The contact force $F$ is :
\begin{equation}
F=N \frac{E^{'}}{2}\sqrt{\pi R\frac{ \sigma^3}{8}} \exp\left(-\frac{2h}{\sigma} \right)
\end{equation}
where $E^{'}$ is the composite elastic modulus et $h$ the distance between the flat surface and the average plane of the summits.

The ratio of force over the contact area gives the average contact pressure:
\begin{equation}
p_m= \frac{F}{A}=\frac{E^{'}}{2}\sqrt{\frac{\sigma}{2 \pi R}} 
\end{equation} 

The elastic energy stored in the contact is:
\begin{equation}
W_{el} = \int_h^{+\infty}F\left(z\right)dz = \frac{\sigma}{2}F\left(h\right)
\end{equation}

\subsection{Plasticity and adhesion}

As proposed by Greenwood and Williamson, if the ratio of the average pressure to the hardness $H$ of the softer material, the amount of plastic deformation of the asperities will increase. This is described by the plasticity index:
\begin{equation}
\psi=\frac{E^{'}}{2H}\sqrt{\frac{\sigma}{R}}=\frac{p_m}{H}\sqrt{2\pi}\approx 2.5 \frac{p_m}{H}
\end{equation}
Generally speaking, plastic deformation will tend to flatten asperities \cite{Ghaednia2017} and make the surface smoother (higher $R$ and lower $\sigma$). Thus the value of $\psi$ will decrease as well as plastic deformation of the surfaces. The hardness is not an intrinsic property of materials and it depends on the contact geometry and work-hardening behavior of the material \cite{Pintaude2023}. A modified version of the plasticity index based on the material yield strength $S_y$ is used \cite{Ghaednia2017}:
\begin{equation}
\psi=\frac{E^{'}}{2CS_y}\sqrt{\frac{\sigma}{R}}
\end{equation}
where $C$ is a proportionality coefficient that is close to 2.8 for limited plastic deformation. The hardness is thus $H=C\times S_y$.

The energy $W_{ad}$ needed to separate the flat and the rough surface is supposed to be equal to the work of adhesion per unit area $w_{ad}$ of the materials times the real contact area $A$ (eq. \ref{eq:A}). Here it is assumed that adhesion does not modify the contact area $A$ due to elastic deformation. The risk of adhesion between the counter-bodies is high if the adhesive energy is higher than the elastic energy stored in the materials. The ratio $\kappa$ of these two energies is :
\begin{equation}
\frac{W_{ad}}{W_{el}}=\kappa=\frac{w_{ad}}{E^{'}}\sqrt{\frac{R}{\sigma^3}}\sqrt{32 \pi}
\end{equation}
This ratio is similar to the adhesive criteria introduced by Gras \cite{Gras2008}.
According to Gras, there will be adhesion if $\kappa > 1$. The adhesion will on the other hand make the surfaces rougher (increase of $\sigma$) due to material transfer or debris formation \cite{Milanese2019, Zhang2022}. Thus is will tend to reduce the value of $\kappa$ and the effect of adhesion.

\subsection{Evolution model}

The model is based on the pioneering work of Nosonovsky \cite{Nosonovsky2010}. He made the assumption that the rate of variation of an arbitrary roughness parameter $Rs$ is linearly dependent on a plastic wear term (negative and proportional to $Rs$) and of an adhesive wear term (positive and inversely proportional to $R$). He obtained a differential equation that he numerically solved. He assumed that the wear was only due to these two phenomena.

In line with the work of Nosonovsky \cite{Nosonovsky2010}, the plastic deformation (controlled by $\psi$) will decrease the roughness height with the sliding distance and the adhesion (controlled by $\kappa$) will increase the roughness height. Using the derivative of the standard deviation $Sq$ of the rough surface with respect to the sliding distance, it is found:

\begin{equation}
\dot{Sq}=-a \psi + b \kappa
\end{equation}
where $a$ and $b$ are positive phenomenological coefficients. It is necessary to express the $\psi$ and $\kappa$ as a function of $Sq$. According to Whitehouse and Archard \cite{Whitehouse1970}, the summit properties can be expressed as a function of the standard deviation and the correlation length of rough profiles. We assumed that their relations can be extended to surface parameters $Sq$ and $Sal$, the correlation length of the rough surface. Thus, $\sigma \propto Sq$ and $R \propto \frac{Sal^2}{Sq}$. Finally, the following ordinary differential equation is found:
\begin{equation}\label{eq:ev_wear}
\dot{Sq}=-a \frac{E^{'}}{2 CSy Sal}Sq +b \frac{w_{ad}Sal}{E^{'}}Sq^{-2}
\end{equation}

It is very similar to the equation proposed by Nosonovsky \cite{Nosonovsky2010} except for the exponent of the adhesive term that is -1 for Nosonosky and -2 here. For simplicity and to be able to derive an analytical solution, it is assumed that all the parameters but $Sq$ are constant. This assumption will be discussed in the last section. The differential equation can thus be reduced to:

\begin{equation}
\dot{Sq}=-\alpha Sq + \beta Sq^{-2}
\end{equation}
where $\alpha$ is the plasticity wear term and $\beta$ the adhesive wear term. This equation has an analytical solution:
\begin{equation}
Sq=\left[\left(Sq_0^3-\frac{\beta}{\alpha}\right)e^{-3\alpha s}+ \frac{\beta}{\alpha}\right]^{1/3}
\end{equation}
where $s$ is the sliding distance and $Sq_0$ is the initial roughness. 

This solution shows a characteristic distance $\tau$ controlled by the plasticity wear term:
\begin{equation}\label{eq:tau}
\tau = \frac{1}{3\alpha} = \frac{2CSy Sal}{3aE^{'}}
\end{equation}

There exists a roughness height limit  $Sq_\infty$ that is independent of the initial roughness:
\begin{equation}\label{eq:Sq_inf}
Sq_\infty =  \left(\frac{\beta}{\alpha}\right)^{1/3} = \left(\frac{2 b w_{ad} CSy Sal^2}{a {E^{'}}^2}\right)^{1/3}
\end{equation}

The solution can be rewritten:
\begin{equation}
Sq=\left[\left(Sq_0^3-Sq_\infty^3 \right)e^{-\frac{s}{\tau}}+ Sq_\infty^3\right]^{1/3}
\end{equation}

The existence of a roughness height limit  $Sq_\infty$ independent of the initial roughness height is consistent with experimental findings \cite{Kumar2002, Bonny2010, PhamBa2021}.

\section{Results}

\subsection{Comparison to experimental results}

Pham-Ba and Molinari \cite{PhamBa2021} did dry sliding tests of polished SiO$_2$ balls on SiO$_2$ disks. Three levels of surface roughness were initially used for the disks. The surface roughness in the wear track was measured at different instants during the 5 h sliding tests. The test conditions were kept constant. They reported the value of the arithmetic roughness height $Sa$ in their paper. In the case of Gaussian surfaces, $Sa \propto Sq$. We will assume that this is verified to be able to use the model based on $Sq$. They found that the three tests, with different initial roughness heights, converged to a unique $Sa_\infty \approx 0.65 \mu$m.

\begin{figure}
\includegraphics[scale = 1]{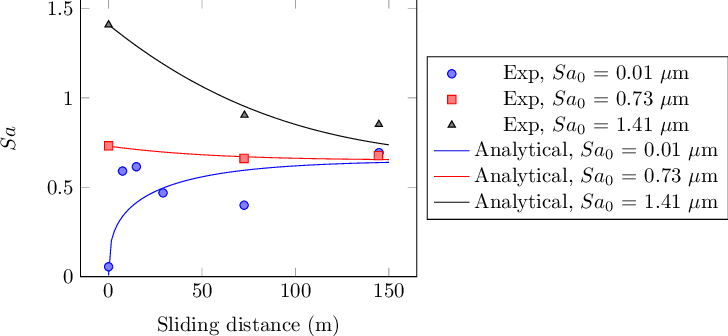}
\caption{Comparison to experimental data \cite{PhamBa2021} with $\tau$ = 50 m and $Sa_\infty$ = 0.65 $\mu$m (it corresponds to $\alpha$ = 0.0666 m$^{-1}$ and $\beta$ = 1.83 $\times$ 10$^{-21}$ m$^2$)}
\label{fig:comp}
\end{figure}

The comparison of the evolution of $Sa$ with the sliding distance is presented in figure \ref{fig:comp} with the characteristic distance $\tau$ = 50 m, obtained by curve fitting. For the initially two roughest surfaces, the $Sa$ value decreases asymptotically to the limit value, as observed in several works \cite{Sugimura1986, Jeng2000, Bonny2010}. The proposed model is able to capture this evolution in a way very close to the experimental points. For the polished disk, the experiments show a rapid increase of the roughness height and then some non-monotonous variations to the final limit value $Sa_\infty$. The model reproduces the rapid evolution of $Sa$ and the convergence to the limit value but it cannot capture the non-monotonous evolution. 

With only two parameters $\tau$ and $Sa_\infty$, the model can correctly reproduce the evolution of the surface roughness during sliding. In the case of the initially polished disk, some intermediate changes are not captured. It can be due to wear debris with a size bigger than the roughness height. Debris are not considered in the model. The addition of the adhesive contribution in the evolution model makes it possible to simulate roughness increase with time which was not possible with former models \cite{Sugimura1986, Jeng2000}. An interesting result is that $\tau$ is the same in the three tests carried out by Pham-Ba and Molinari \cite{PhamBa2021}, indicating that the plastic wear coefficient $\alpha$ is an intrinsic property of the sliding surfaces.

\subsection{Influence of the model parameters}

The plastic deformations of the asperity tend to reduce the roughness height at a rate proportional to $Sq$ with a coefficient:
\begin{equation}
\alpha=a \frac{E^{'}}{2CSy Sal}
\end{equation}
On the other hand, the adhesion tends to increase the roughness height at a rate proportional to $Sq^{-2}$ with a coefficient:
\begin{equation}
\beta=b \frac{w_{ad}Sal}{E^{'}}
\end{equation}
The steady-state roughness height is reached when these tow terms are exactly equal and vanish. It means that the roughness height limit is the result of the competition between the plastic deformation and adhesion. The speed at which the steady wear regime is reached is controlled by $\alpha$, the plasticity wear term.

In this section, the evolution of the surface roughness during the running-in period when $\alpha$ and $\beta$ are varied is presented and discussed. The values of dimensionless parameters $\alpha \cdot Sq_0$ and $\beta/Sq_0^2$ were chosen in a range consistent with the experiments of Pham-Ba and Molinari \cite{PhamBa2021}.

\begin{figure}
\includegraphics[scale = 1]{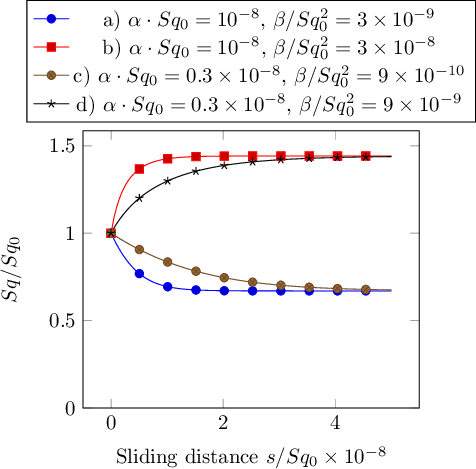}
\caption{Evolution of the surface roughness during the running-in period when the plasticity wear term $\alpha$ and the adhesive wear term $\beta$ are varied}
\label{fig:param}
\end{figure}

Figure \ref{fig:param} presents the evolution of the surface roughness height for four sets of values of $\alpha$ and $\beta$. In the first case a), the ratio of $\beta$ over $\alpha$ is lower than $Sq_0^3$ thus the final roughness is lower than $Sq_0$. For case b), $\alpha$ is kept constant, and $\beta$ is raised so the roughness height increases to a value higher than $Sq_0$. As $\alpha$ does not vary, the distance needed to reach the steady state remains the same. For case c) (respectively d), the values of case a) (respectively b) are decreased in the same ratio. Because of that, the final roughness height is the same but the distance needed for steady state is higher.

\subsection{Preponderant plasticity}

\begin{figure}
\includegraphics[scale = 1]{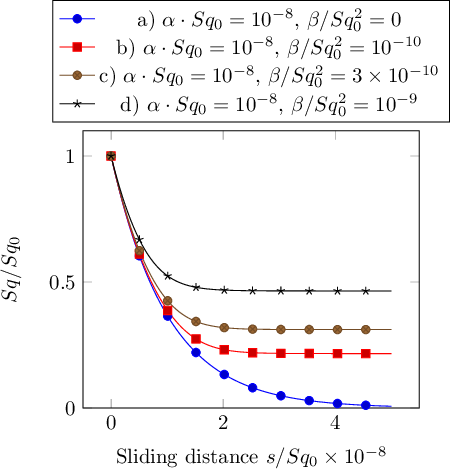}
\caption{Evolution of the surface roughness during the running-in period when the plasticity wear term $\alpha$ is constant and the adhesive wear term $\beta$ is varied}
\label{fig:plasticity}
\end{figure}

If there is no adhesion, meaning that $\beta$ = 0, the steady-state roughness will be zero. In this case, the plasticity is preponderant. It corresponds to an ideal polishing process. This situation corresponds to case a) in figure \ref{fig:plasticity}. When a small amount of adhesion is added by increasing $\beta$, a limit value $Sq_\infty$ of the roughness different from zero is reached. The value of $Sq_\infty$ increases with $\beta$ as shown in figure \ref{fig:plasticity}, cases b) to d).

\subsection{Preponderant adhesion}

\begin{figure}
\includegraphics[scale = 1]{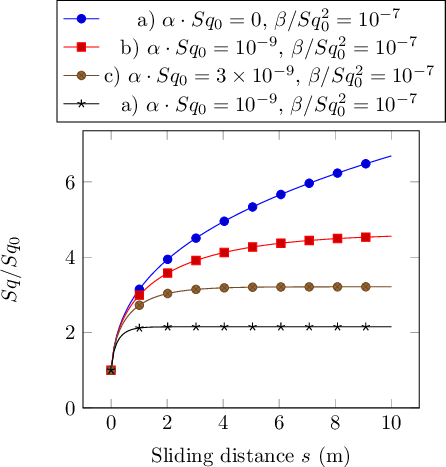}
\caption{Evolution of the surface roughness during the running-in period when the plasticity wear term $\alpha$ is varied and the adhesive wear term $\beta$ is constant}
\label{fig:adhesion}
\end{figure}

If there is no plastic deformation ($\alpha = 0$), the adhesive wear is preponderant. In this case, the solution of the differential equation is different:
\begin{equation}
Sq=\left(3\beta s+Sq_0^3\right)^{1/3}
\end{equation}
The analytical solution shows that the roughness height can only increase with the sliding distance $s$. In addition, the roughness height $Sq$ has no limit. It can be compared to an extreme adhesive wear situation such as galling. The evolution of the roughness height in such a situation is presented in figure \ref{fig:adhesion}, case a). When $\alpha$ is gradually increased, it is possible to reach a steady state regime, cases b) to d). The value of $Sq_\infty$ as well as the sliding distance needed to reach the steady state decreases when $\alpha$ is higher. 

\section{Discussion}

The sliding distance $\tau$ needed to reach the steady state is proportional to the hardness $CSy$ and the correlation length $Sal$ of the rough surface (see Eq. \ref{eq:tau}). Increasing the hardness of the material will limit plastic deformation. In addition, increasing the correlation length will increase the radius $R$ of the asperities and thus limit plastic deformation. On the other hand, a stiffer set of materials with higher $E^{'}$ will promote plastic deformation and reduce the value of $\tau$.

The value of the steady-state roughness height $Sq_\infty$ (Eq. \ref{eq:Sq_inf}) evolves in the same way as $\tau$ when $CSy$, $Sal$ or $E^{'}$ are varied. The work of adhesion $w_{ad}$ of the material will also increase the final roughness height in the steady state because it will promote adhesive wear. The increase of the steady-state roughness height with the hardness of the material is consistent with the findings of Bonny et al. \cite{Bonny2010}. As previously discussed, $Sq_\infty$ results from the competition between abrasive wear and adhesive wear. Aghababaei and al. \cite{Aghababaei2022} have already shown that surface roughness is controlled by the wear mechanisms involved during the rubbing process.

These two parameters are in addition influenced by two phenomenological coefficients $a$ and $b$. Using the data of Pham-Ba and Molinari \cite{PhamBa2021} and assuming a hardness of $CSy$ = 6 GPa and $Sal$ = 10 $\mu$m for the amorphous silica, it is found that $a \approx 6\times 10^{-9}$ and $b \approx 10^{-5}$. If the plasticity index $\phi$ and the adhesion index $\kappa$ are equal to one, $a$ and $b$ indicate the variation of $Sq$ in meter per unit sliding distance. During the sliding, debris due to adhesion will be generated and increase the roughness height. The amount of variation is proportional to $b$ and the adhesion index $\kappa$. At the same time, the newly generated roughness peaks will be flattened by plastic deformation, decreasing thus $Sq$. The amount of decrease in roughness is proportional to $a$ and the plasticity index. In the work of Pham-Ba and Molinari \cite{PhamBa2021}, the values of $a$a and $b$ indicate that adhesion is almost three orders of magnitude more efficient in modifying roughness height than plastic deformation. However, the value of $\kappa$ is quickly reduced due to the dependence on $Sq^{-2}$ limiting thus the adhesion when the roughness height increases. The values of $a$ and $b$ is probably related to the critical junction size that determines if a contacting asperity will be plastically deformed or will generate a debris \cite{Aghababaei2016}. In the experimental studies used for comparisons (\cite{Bonny2010,PhamBa2021}), it can be expected that increasing the load will modify the values of $a$ and $b$ because the number of contacting asperities will grow with the load. Thus more asperities will be subjected to wear.

In the present work, it has been assumed that all the parameters except the roughness $Sq$ are constant in Eq. \ref{eq:ev_wear} to obtain an analytical solution. As previously discussed, $a$ and $b$ can vary. More generally, in real situations, all the other equation parameters in Eq. \ref{eq:ev_wear} can change during running-in making the resolution more difficult. However, if it is possible to estimate these parameters at each time, the evolution equation can be solved numerically. Then it is possible to consider the effect of other physical phenomena and obtain possible non-monotonic evolution of $Sq$.

For example, the change in material properties due to plastic strain can be considered by modifying $C$. The temperature rise can promote oxidation that will affect the value of the work of adhesion $w_{ad}$.

If one of the surfaces is assumed to be flat and rigid, the correlation length $Sal$ can be reasonably considered constant. However in real situations when two rough surfaces interact, the correlation length will change with time as shown in the work of Minet et al. \cite{Minet2010}. $Sq$ and $Sal$ describe respectively the height and characteristic size of asperities. Their evolution is certainly coupled. It means that a second evolution equation for $Sal$ is necessary but this is out of the scope of the present work. In the case where the rigid surface is a machining or polishing tool, the $Sal$ of the worn surface will converge to a value close to the one of the hard tool surface, as described by Whitehouse \cite{Whitehouse2001}. 

In the case of a hard polishing tool, the roughness height of the soft surface will not converge to zero even if there is no adhesion (which is not realistic) as described in the section on preponderant plasticity. Indeed, at the end of the polishing process, the worn surface and the tool will be commensurate due to the plowing of the tool in the soft material. The combined $Sq$ of both surfaces will tend to zero making the value $\dot{Sq}$ vanishing in Eq. \ref{eq:ev_wear}. The final $Sq$ of the worn surface will be close to the one of the tool but not zero.

The presence of wear debris in the contact is neglected in the model. The debris can be considered by using a contact model that takes into account spherical particles between the rubbing surfaces, as done by Horng et al.\cite{Horng2021}. In addition, it is necessary to have a model giving the size and quantity of debris generated during sliding, as, models based on the the critical junction size concept \cite{Aghababaei2016}. For initially smooth surfaces, the generation of debris of size bigger than the roughness height can significantly affect the evolution of $Sq$. It can explain the non-monotonic evolution of the roughness height of smooth surfaces in the experimental work of Pham-Ba and Molinari \cite{PhamBa2021}. The debris themselves can experience plastic deformation (flattening \cite{PhamBa2021}) and adhesion during sliding. Considering debris is a necessary but complicated step for future work.

\section{Conclusion}

In this paper, an analytical expression of the roughness height evolution during the running-in period of sliding friction is proposed. It is based on the model of evolution proposed by Nosonovsky \cite{Nosonovsky2010}. The originality of the present work is to use the plasticity index and the adhesion index as the coefficients of the evolution law, which is thus physically based. The model has been successfully compared to experiments where the roughness can either decrease or increase during the running-in period depending on the initial roughness height of the surfaces. It is shown that the final steady-state roughness height is the result of the competition between the adhesion to the plastic deformation (ratio of the adhesive wear coefficient to the plastic wear coefficient). The distance needed to reach the stable regime is inversely proportional to the plastic wear coefficient. If there is no adhesion, the final roughness height is zero. It is an ideal polishing process. On the other hand, if there is no plastic deformation, the roughness height will increase without any limit, corresponding to a severe adhesive wear regime. 

This model is very simple and relies on strong assumptions but it is able to capture the main trends of the evolution of roughness during running-in. It can thus be a useful tool for a rapid analysis of the effect of physical parameters on the roughness changes during sliding wear.

\section*{Acknowledgments}

This work pertains to the French government program France 2030 (LABEX INTERACTIFS, reference ANR-11-LABX-0017-01 and EUR INTREE, reference ANR-18-EURE-0010).


\bibliography{biblio/ref}

\end{document}